# Turbulence closure in the light of phase transition

**Mohammed A. Azim**[a,b]

Department of Mechanical Engineering, Bangladesh University of Engineering and Technology, Dhaka-1000, Bangladesh, email: azim@me.buet.ac.bd

**Abstract**

In this study, new turbulence closure equations are derived in the light of turbulence as a continuous phase transition phenomenon. Closed-form Reynolds averaged Navier-Stokes equations due to those closure equations are solved numerically by treating turbulent viscosity as a tensor, unlike the eddy viscosity, for a plane turbulent jet. Overall agreement between the obtained results and the existing literature for the jet flow demonstrates the effectiveness of the new closure equations. Besides, turbulent stresses as a function of the normalized mean velocity exhibit their odd and even symmetries in the flow, which are manifestations of the free energy symmetry of continuous phase transition.

## 1 Introduction

Turbulent flow is characterized by a hierarchy of scales of fluctuations ranging from production to dissipation. The dynamics of turbulent flow is governed by the Navier-Stokes equations, whose solution resolving all those scales at high Reynolds number in complex geometries is unlikely to be attainable in the foreseeable future. However, Reynolds averaging and filtering are the two methods that have been used to transform the Navier-Stokes equations so that the small-scale turbulent fluctuations do not have to be directly simulated. Both methods introduce additional terms in the governing equations due to the correlations of fluctuating quantities (i.e., turbulent stresses) that need to be modeled in order to achieve closure. Turbulence modeling requires that turbulent stresses in the Reynolds averaged Navier-Stokes (RANS) equations be appropriately modeled [1]. Boussinesq hypothesis [2] is the core of all turbulence models that relates the Reynolds stresses to the mean velocity gradients as

$$-\overline{u_i'u_j'} = \nu_T\left(\partial\bar{u}_i/\partial x_j + \partial\bar{u}_j/\partial x_i\right) \tag{1}$$

for an incompressible flow where $\nu_T$ is the eddy viscosity that requires further modeling as, for example, $\nu_T = \ell_m^2\left(\partial\bar{u}/\partial y\right)$ after Prandtl [3], $\nu_T = C_\mu \ell_m \sqrt{k}$ after Kolmogorov [4] and Prandtl [5], $\nu_T = C_\mu k^2/\varepsilon$ after Launder and Spalding [6], where $\ell_m$ is the mixing length (closely related to the length scale), $k$ the turbulent kinetic energy and $\varepsilon$ the dissipation rate of $k$.

a Former Professor, b Present Address: Nirvana Residential Area (3rd flr, Bldg 4), 30/31 College Road, Chattogram-4203, Bangladesh.

The present study aims to derive the closure equations by using the features of continuous (also called second-order) phase transition against the order parameter. The concept of an order parameter in phase transitions has been widely generalized as when an arbitrary system crosses an instability point, only a few variables become relevant and serve as an order parameter, which may be a single variable or a set of coupled variables that distinguish the ordered from the disordered phase. However, an averaged velocity may be a suitable control parameter for a fluid dynamical system, which, if adequately tuned, can undergo great qualitative changes in its macroscopic properties and, consequently, in microscopic properties. Landau [7] observed during a second-order phase transition that the system becomes more orderly as it cools through the transition temperature and proposed to measure the orderliness by a field variable called the order parameter. Goldenfeld [8] showed that the rise in temperature in metals is similar to the increase in velocity in fluid flows; the former causes magnetization at critical (transition) point, and the latter turbulence. The above observations invoke a strong analogy between turbulization and magnetization at critical points [7-10]. Recently, many researchers have demonstrated experimentally and numerically by resolving the time and length scales sufficiently close to the critical point that all turbulent flows are closely analogous to the continuous phase transition (CPT) phenomena [11-14]. The free energy of a system and the kinetic energy of a turbulent flow are analogous because both near their critical points obey the power-law scaling of the correlation functions [15]. Further, the correlation functions for turbulence are strictly short-ranged [16] and become maximum [17] near the critical point. In CPTs, the derivative $\partial F/\partial \Phi$ is continuous and $\partial^2 F/\partial \Phi^2$ is discontinuous at the point of transition, i.e., $\partial F/\partial \Phi = 0$ and $\partial^2 F/\partial \Phi^2 < 0$ where $F$ is the free energy of the system and $\Phi$ the order parameter.

## 2 Closure equations due to phase transition

Navier-Stokes equations represent the collective motion of fluid particles. These equations of motion for incompressible fluid are

$$\frac{\partial \bar{u}_i}{\partial t} + \bar{u}_j \frac{\partial \bar{u}_i}{\partial x_j} = -\frac{\partial}{\partial x_j}\left(\frac{\bar{p}}{\rho}\delta_{ij} + \overline{u'_i u'_j}\right) + \nu \frac{\partial^2 \bar{u}_i}{\partial x_j \partial x_j}. \tag{2}$$

In collective motion [18], the most naturally (but not necessarily) chosen order parameter is the predominant velocity $\bar{u}$. The concept of moving equilibrium [19] states that the rate of change of each component of the Reynolds stress $\overline{u'_i u'_j}$ is proportional to the rate of change of turbulence kinetic energy $k$, that is, $\overline{u'_i u'_j}$ represents the free energy of turbulence.

In view of relating the features of phase transition to the dynamics of turbulence, an implicit function $f\left(\bar{u}_i, R_{ij}\right) = 0$ for $R_{ij} = \overline{u'_i u'_j}$ is written by taking divergence of Eq. (2) and subsequently applying the Gauss divergence theorem as

$$f\left(\bar{u}_i, R_{ij}\right) = \frac{\partial}{\partial x_j}\left(\bar{u}_i \bar{u}_j + \frac{\bar{p}}{\rho}\delta_{ij} + \overline{u'_i u'_j}\right). \tag{3}$$

The differential form of the implicit function is $\left(\partial f/\partial \bar{u}_i\right)d\bar{u}_i + \left(\partial f/\partial R_{ij}\right)dR_{ij} = 0$. At phase transition, the fluctuation correlations $\overline{u'_i u'_j}$ become maximum [17], that is, $dR_{ij}$ become zero leading to $\partial f/\partial \bar{u}_i = 0$ because $d\bar{u}_i(x_j) = 0$ is unrealistic. Equation $\partial f/\partial \bar{u} = 0$ includes the first derivative of the free energy with respect to the order parameter $\bar{u}$ that expresses the state of turbulence in terms of free energy equilibrium and thus, combines the dynamics of turbulence with that of CPT as

$$\frac{\partial}{\partial \bar{u}}\frac{\partial}{\partial x_j}\left(\bar{u}_i \bar{u}_j + \frac{\bar{p}}{\rho}\delta_{ij} + \overline{u'_i u'_j}\right) = 0 . \tag{4}$$

Equation (4) can be written using an entity $\left(f_{(i)(j)}(\bar{u}) + f_1\right)\partial \bar{u}_j/\partial x_j = 0$ due to the continuity equation as

$$\frac{\partial}{\partial \bar{u}}\frac{\partial}{\partial x_j}\left(\bar{u}_i \bar{u}_j + \frac{\bar{p}}{\rho}\delta_{ij} + \overline{u'_i u'_j}\right) = \left(f_{(i)(j)} + f_1\right)\frac{\partial \bar{u}_j}{\partial x_j} \tag{5}$$

where $f_1$ is an arbitrary function. The indices (*i*) and (*j*) in the parentheses, hereafter, indicate that they are not summed over.

Equation (5) for the Reynolds stress $\overline{u'u'}$, may be written as

$$\frac{\partial}{\partial \bar{u}}\frac{\partial}{\partial x}\left(\bar{u}\bar{u} + \frac{\bar{p}}{\rho} + \overline{u'u'}\right) = \left(f_{11} + f_1\right)\frac{\partial \bar{u}}{\partial x} \tag{6}$$

and which can be written successively as

$$\frac{\partial}{\partial \bar{u}}\frac{\partial}{\partial y}\left(\bar{u}\bar{u} + \frac{\bar{p}}{\rho} + \overline{u'u'}\right)\frac{\partial y}{\partial x} = \left(f_{11} + f_1\right)\frac{\partial \bar{u}}{\partial y}\frac{\partial y}{\partial x} \tag{7}$$

$$\frac{\partial}{\partial \bar{u}}\frac{\partial}{\partial y}\left(\bar{u}\bar{u} + \frac{\bar{p}}{\rho} + \overline{u'u'}\right) = \left(f_{11} + f_1\right)\frac{\partial \bar{u}}{\partial y} \tag{8}$$

$$\frac{\partial}{\partial \bar{u}}\frac{\partial}{\partial \bar{u}}\left(\bar{u}\bar{u} + \frac{\bar{p}}{\rho} + \overline{u'u'}\right) = f_{11}(\bar{u}) + f_1 . \tag{9}$$

Assuming $f_1 = 0$, Eq. (9) becomes

$$\frac{\partial}{\partial \bar{u}}\frac{\partial}{\partial \bar{u}}\left(\bar{u}\bar{u} + \frac{\bar{p}}{\rho} + \overline{u'u'}\right) = f_{11}(\bar{u}) . \tag{10}$$

For stability reason, Eq. (10) is written as

$$\frac{\partial}{\partial \bar{u}}\frac{\partial}{\partial \bar{u}}\left(\overline{u'u'} + \frac{\bar{p}}{\rho}\right) = f_{11}(\bar{u}) - 2 . \tag{11}$$

Considering $f_1 = \frac{\partial}{\partial \bar{u}}\frac{\partial}{\partial \bar{u}}\left(\frac{\bar{p}}{\rho}\right) + 2$, Eq. (9) may be written as

$$\frac{\partial}{\partial \bar{u}}\frac{\partial}{\partial \bar{u}}\left(\overline{u'u'}\right)= f_{11}(\bar{u})\,. \tag{12}$$

Equations like (11), may be obtained from Eq. (5) for the Reynolds stresses $\overline{v'v'}$ and $\overline{u'v'}$. Thus, the equations for Reynolds stresses may be read by integrating across the shear layer as

$$\overline{u'u'} = -\frac{\bar{p}}{\rho}+\iint\left(f_{11}-2\right)d\bar{u}d\bar{u} \tag{13a}$$

$$\overline{v'v'} = -\frac{\bar{p}}{\rho}+\iint\left(f_{22}-2\frac{\partial \bar{v}}{\partial \bar{u}}\right)d\bar{u}d\bar{v} \tag{13b}$$

$$\overline{u'v'} = \iint\left(f_{12}-1\right)d\bar{u}d\bar{v}\,, \tag{13c}$$

and the function $f_{ij}(\bar{u})$ for each stress may be obtained following Eq. (12) as

$$f_{11}(\bar{u}) = \frac{\partial}{\partial \bar{u}}\frac{\partial}{\partial \bar{u}}\left(\overline{u'u'}\right),\; f_{22}(\bar{u}) = \frac{\partial}{\partial \bar{u}}\frac{\partial}{\partial \bar{v}}\left(\overline{v'v'}\right),\text{ and }\; f_{12}(\bar{u}) = \frac{\partial}{\partial \bar{u}}\frac{\partial}{\partial \bar{v}}\left(\overline{u'v'}\right). \tag{14}$$

Reduction of Eq. (8) to Eq. (9) involves the transformation of a differential operator, that is, $\frac{\partial}{\partial \bar{u}}\frac{\partial}{\partial y}(\;)$ on the space coordinate $y(0,\infty)$ is transformed to $\frac{\partial}{\partial \bar{u}}\frac{\partial}{\partial \bar{u}}(\;)$ on the space coordinate $y'(-\infty,\infty)\equiv 2y'(0,\infty)$. This implies the equations for Reynolds stresses are in $y'$ coordinate. In the Gaussian profile of streamwise velocity, $\bar{u} = u_c\, exp\left(-0\cdot 693y^2/b^2\right)$ with $b=y_{1/2}$ the jet half-width, $\bar{u}$-coordinate corresponds to $y^2$. Hence, $2y'^2 = y^2$, that is, $y = \sqrt{2}\,y'$.

Let the free energy of turbulence be defined as

$$F_{ij} = \overline{u'_i u'_j} - \overline{u'_{(i)}u'_{(j)}}\left(\bar{u}/u_c\right)\delta_{ij} = \overline{u'_i u'_j} - b_{(i)}\bar{u}^2\left(\bar{u}/u_c\right)\delta_{ij} \tag{15}$$

where the first term on the right-hand side is the energy that enters into the turbulence by the interaction of fluctuations and mean field shear, and the second term is the energy that leaves the turbulence due to the coupling between turbulent fluctuations and order parameter, and all these energies finally dissipate as heat. Landau-free energy is a thermodynamic potential of a system despite being expressed as a function of the order parameter and its conjugate external field. In Eq. (15), $\overline{u'_{(i)}u'_{(j)}}\,\delta_{ij}$ is replaced by $b_{(i)}\bar{u}^2\delta_{ij}$ using the order of magnitude reasoning.

Equation (13) may be read in tensor notation as

$$\overline{u'_i u'_j} = -\frac{\bar{p}}{\rho}\delta_{ij} + \iint\left(f_{ij} - \frac{\partial}{\partial \Phi}\frac{\partial}{\partial \bar{u}_j}\left(\Phi \bar{u}_{(j)}\right)\right)d\Phi d\bar{u}_{(j)} \tag{16}$$

where $\Phi\left(=\bar{u}\right)$ is the order parameter. On replacing $\overline{u'_i u'_j}$ (free energy without additional term) by $\overline{u'_i u'_j} - b_{(i)}\bar{u}^2\Phi_n\,\delta_{ij}$ (free energy with additional term), Eq. (16) becomes

$$\overline{u_i'u_j'} = b_{(i)}\bar{u}^2\Phi_n\,\delta_{ij} - \frac{\bar{p}}{\rho}\delta_{ij} + \iint\left(f_{ij} - \frac{\partial}{\partial\Phi}\frac{\partial}{\partial\bar{u}_j}\left(\Phi\bar{u}_{(j)}\right)\right)d\Phi d\bar{u}_{(j)} \tag{17}$$

where $b_i$ is a constant tensor and $\Phi_n = \bar{u}/u_c$. Equation (17) provides the closure equations as

$$\overline{u_i'u_j'} = C_{(i)(j)}\left[b_{(i)}\bar{u}^2\Phi_n\delta_{ij} - \frac{\bar{p}}{\rho}\delta_{ij} + \iint\left(f_{ij} - \frac{\partial}{\partial\Phi}\frac{\partial}{\partial\bar{u}_j}\left(\Phi\bar{u}_{(j)}\right)\right)d\Phi d\bar{u}_{(j)}\right] \tag{18}$$

where $C_{ij}$ are the scaling constants to balance between the turbulent and mean quantities.

## 3 Application to a plane jet flow

Considering turbulence as a continuous phase transition phenomenon, the derived closure equations are applied to unclosed RANS equations for a plane jet as a test case. Plane jets are free shear flows where fluids coming out from a two-dimensional (2D) orifice mix with the ambient fluid and develop through three successive distinct regions, namely initial, intermediate, and developed regions (Fig. 1). Initial region (at $x/h \leq 6$) is characterized by the presence of a potential core of uniform axial velocity that is in laminar state, intermediate region by the transition state that is triggered by the fluctuations of local flow properties and developed region by the turbulent state. Literature [20] shows a large scatter of the downstream distance at which the jet becomes fully developed, ranging from *5h to 25h*, depending on its initial conditions (e.g., Reynolds number at exit and nozzle geometry). This jet grows nonlinearly in the developing (initial and intermediate) region and linearly in the developed (self-similar) region, e.g., the growth of jet half-width. The governing equations and their initial and boundary conditions, as well as the determination of the constants $C_{ij}$ are presented in this section.

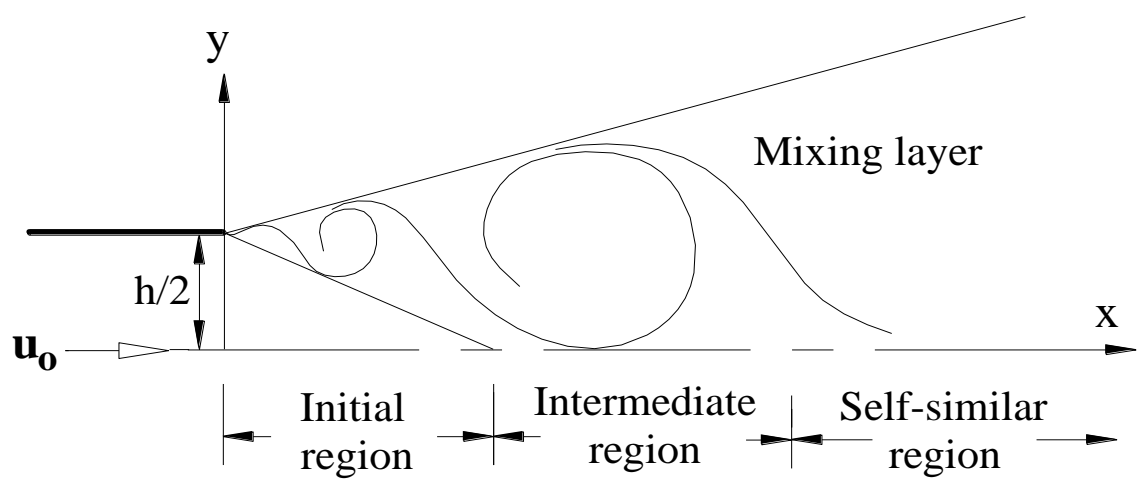

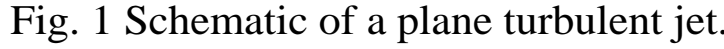
Fig. 1 Schematic of a plane turbulent jet.

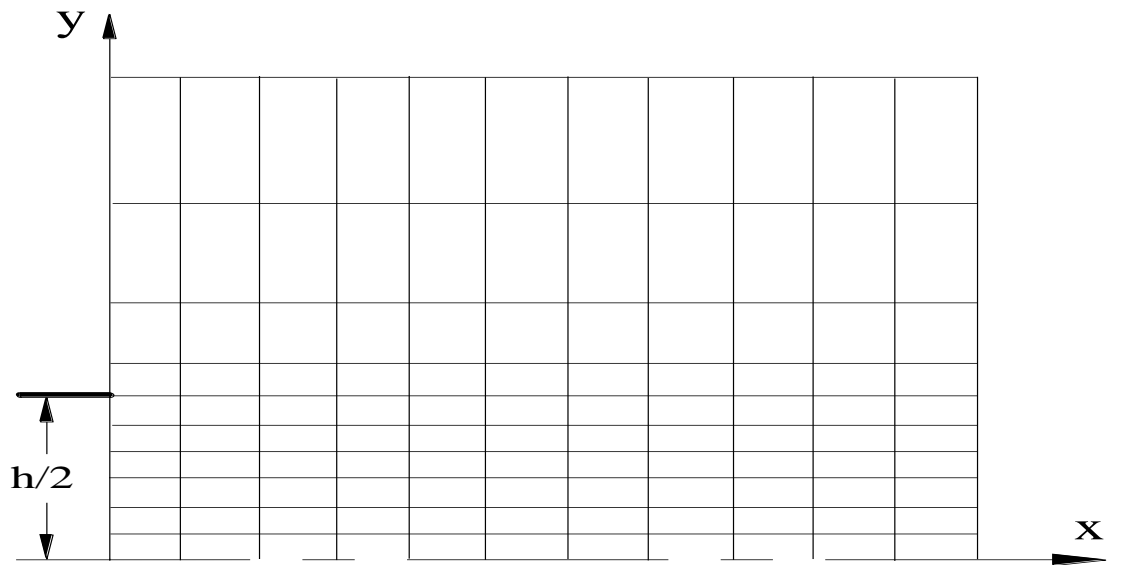

Fig. 2 Computational grids for a plane turbulent jet.

### 3.1 Governing equations

Continuity and RANS equations describing a steady plane turbulent jet flow for incompressible fluid are

$$\frac{\partial\bar{u}_i}{\partial x_i} = 0 \tag{19}$$

$$\bar{u}_j \frac{\partial \bar{u}_i}{\partial x_j} = -\frac{1}{\rho}\frac{\partial \bar{p}}{\partial x_i} + \frac{\partial}{\partial x_j}\left( \nu \frac{\partial \bar{u}_i}{\partial x_j} - \overline{u'_i u'_j} \right) \tag{20}$$

where $-\overline{u'_i u'_j} = \nu_{T(i)(j)}\left( \partial \bar{u}_i / \partial x_j + \partial \bar{u}_j / \partial x_i \right)$ is the turbulent stress-strain rate relation and $\nu_{Tij}$ the turbulent viscosity tensor. Equations (18)-(20) are in closed-form and need to be solved to extract the mean and turbulence quantities of the flow.

### 3.2 Initial and boundary conditions

In free jet flow, the initial conditions are $\bar{u}(0, y \le 0 \cdot 5h) = u_o$, $\bar{u}(0, y > 0 \cdot 5h) = u_e$ and $\bar{v}(0, y) = 0$ where $h$ is the orifice height, $u_o$ the uniform jet exit velocity and $u_e$ the constant velocity beyond the jet outer edge. The boundary conditions are: $\phi$ (general flow variable) attains ambient conditions at the jet outer edge $y_e$, $\partial\phi / \partial x = 0$ at the outflow, and $\partial\phi / \partial y = 0$ at the symmetry axis except $\bar{v} = 0$ and $\overline{u'v'} = 0$.

### 3.3 Determination of $C_{ij}$

The concept of moving equilibrium [19] may be read in explicit form as

$$\overline{u'_i u'_j} = a_{(i)(j)}\left( \delta_{ij} \pm \varepsilon_{ij\ell} \right) k \tag{21}$$

where $a_{ij}$ are the proportionality constants. Equating Eqs. (18) and (21), one can write

$$C_{(i)(j)}\left[ b_{(i)}\bar{u}^2 \Phi_n \delta_{ij} - \frac{\bar{p}}{\rho}\delta_{ij} + \iint \left( f_{ij} - \frac{\partial}{\partial \Phi}\frac{\partial}{\partial \bar{u}_j}\left( \Phi \bar{u}_{(j)} \right) \right) d\Phi d\bar{u}_{(j)} \right] = a_{(i)(j)}\left( \delta_{ij} \pm \varepsilon_{ij\ell} \right) k \tag{22}$$

which shows that $C_{ij}=a_{ij}$. Using $-\overline{u'v'} / \sqrt{\overline{u'u'}}\sqrt{\overline{v'v'}} \approx 0 \cdot 48$ and $\overline{u'v'} / k \approx 0 \cdot 35$ for turbulent flows, and $\overline{u'u'} \approx 1 \cdot 4\overline{v'v'}$ for the plane jet flow [21] along with Eq. (21), the constants are evaluated as $a_{11}=0{\cdot}862$, $a_{22}=0{\cdot}616$, and $a_{12}=0{\cdot}35$.

### 3.4 Determination of $f_{ij}(\bar{u})$ and closure equations in $(x, y, z)$

The free energy of turbulence $\overline{u'_i u'_j}$ may be expressed by a fourth degree polynomial of normalized velocity $\bar{u}/u_c$ with a view to capture the free energy symmetry of continuous phase transition. Analysis of turbulent shear stress data of plane and circular jets [20,21] show that the coefficient of the fourth-order term is insignificant compared to other coefficients of that polynomial in the initial region of the jet while the same is more significant than or comparable to other coefficients in the intermediate and self-similar regions. Furthermore, the coefficients of the second-order term of the polynomials higher than the fourth degree are found small, which is inconsistent with the order of magnitude estimate of $\overline{u'v'}$ by dimensional analysis. It seems now that $\overline{u'v'}$ is best expressed by a third degree polynomial of $U = \bar{u}/u_c$ in the initial region of the jet and by a fourth degree polynomial beyond this region. Such a polynomial for the turbulent shear stress may be assumed as $-\overline{u'v'}/u_c^2 = C_s\left(U - U^2 + U^3 - U^4\right)$. This shear stress is generally symmetric with respect to the origin (called odd symmetric) while the normal stresses are symmetric with respect to the axis (called even symmetric). Hence, turbulent normal stress function may be obtained by integrating the above polynomial as $\overline{u'u'}/u_c^2 = C_n U\left|1 - \frac{1}{2}U + \frac{1}{3}U^2 - \frac{1}{4}U^3 + \frac{1}{5}U^4 + rest\right|$, which further can be approximated as $\overline{u'u'}/u_c^2 = C_n U\, exp\left(-\alpha U\right)$ where $C_s$, $C_n$ and $\alpha$ are constants. This is because an odd symmetric function, either by differentiation once or by integration, becomes even symmetric and vice-versa.

The function $f_{ij}(\bar{u})$ for the Reynolds stresses is obtained from Eq. (14) using their functional form for the normal stress and the polynomial form for the shear stress. The derived function for the axial normal stress $\overline{u'u'}$ is

$$f_{11}(\bar{u}) = \partial^2\left(\overline{u'u'}\right)/\partial u^2 = C_n u_c \beta\left(-2 + \beta\bar{u}\right)\exp\left(-\beta\bar{u}\right) \tag{23a}$$

where $\beta = \alpha/u_c$, for the transverse normal stress $\overline{v'v'}$ is

$$f_{22}(\bar{u}) = \partial^2\left(\overline{v'v'}\right)/\partial u\partial v = \left(a_{22}/a_{11}\right)\left(\partial\bar{u}/\partial\bar{v}\right)f_{11}(\bar{u}) \tag{23b}$$

and for the shear stress $\overline{u'v'}$ is

$$f_{12}(\bar{u}) = \partial^2\left(\overline{u'v'}\right)/\partial u\partial v = C_s\left(\bar{u}/u_c, \bar{u}^2/u_c^2\right)\partial u/\partial v\,. \tag{23c}$$

Using the polynomial for $\overline{u'v'}$, one obtains $f_{12}(\bar{u}) = C_s\left(\bar{u}/u_c\right)\partial u/\partial v$ for the initial region of the jet and $f_{12}(\bar{u}) = C_s\left(\bar{u}^2/u_c^2\right)\partial u/\partial v$ for areas beyond this region. Here, the highest order term of the polynomial for $\overline{u'v'}$ is used instead of the entire polynomial as this requires known values of the coefficients that might cause ascertaining $f_{12}(\bar{u})$ unlikely. Further, the function $f_{12}(\bar{u})$ is the second derivative of the highest order term of the polynomial, as a result, very closely retains the same features of $f_{12}(\bar{u})$ if were obtained by using the entire polynomial. Above reasons justify the use of the highest order term instead of the entire polynomial in determining $f_{12}(\bar{u})$. Further, the derivative of the continuity equation with respect to $\bar{u}$ results in $\partial\left(\partial v/\partial u\right)/\partial y = 0$ which implies $\partial v/\partial u = f_2\left(x\right)$ or a constant.

Closure equation (18) in $(x,y,z)$ coordinates becomes

$$\overline{u'u'} = C_{11}\left(b_1\bar{u}^3/u_c - \frac{\bar{p}}{\rho} + \iint\left(f_{11} - 2\right)d\bar{u}d\bar{u}\right) \tag{24a}$$

$$\overline{v'v'} = C_{22}\left( b_2 \bar{u}^3 / u_c - \frac{\bar{p}}{\rho} + \iint \left( f_{22} - 2\partial\bar{v}/\partial\bar{u} \right) d\bar{u}\, d\bar{v} \right) \tag{24b}$$

$$\overline{u'v'} = C_{12} \iint \left( f_{12} - 1 \right) d\bar{u}\, d\bar{v}\,. \tag{24c}$$

## 4 Numerical strategies

A Computational Fluid Dynamics (CFD) code is developed for solving 2D steady flow based on finite volume method with rectangular structured grid [22]. SIMPLE (Semi-Implicit Method for Pressure Linked Equations) algorithm of Patankar and Spalding [23] is employed to solve the equations (18)-(20) governing the plane turbulent free jet. Discretization of the governing equations, computational grids, and grid convergence tests are described in this section.

### 4.1 Discretization and computational grids

The convective and diffusive terms of the transport equations for $\phi$ are discretized using the second-order upwind difference scheme and the second-order central difference scheme in the staggered grid, respectively, with cell centers for $\bar{u}$ at (i,J), for $\bar{v}$ at (I,j), for $\bar{p}$, $\overline{u'u'}$ and $\overline{v'v'}$ at (I,J), and for $\overline{u'v'}$ at (i,j). The discretized equations are solved iteratively using the line-by-line tridiagonal matrix algorithm (TDMA) [24]. The momentum and pressure correction equations are provided with a suitable number of sweeps. All variables $\left(\bar{u}, \bar{v}, \bar{p}\right)$ are weighted with low, fixed under-relaxation factors for the stability of the numerical scheme. The flow variables are further updated by a weighted average of their neighbors in each iteration, after being obtained using TDMA. This procedure ensures the stability of the numerical scheme, in addition to under-relaxation factors, by reducing oscillations of the strain rates in turbulent viscosity calculation.

The flow domain is constructed over $40h \times 20h$ in $x \times y$-directions with grids as in Fig. 2. Grids are uniform inside the orifice and varying outside it in $y$-direction, and uniform in $x$-direction as in the figure. The solutions are duly converged with normalized residuals of $3{\cdot}9 \times 10^{-6}$, $2{\cdot}7 \times 10^{-5}$ and $2{\cdot}0 \times 10^{-4}$ for $\bar{u}$, $\bar{v}$ and mass, respectively.

### 4.2 Grid convergence test

The present simulation is performed for a plane turbulent air jet with co-flow, and Reynolds number $Re(=u_o h/\nu) = 3{\cdot}2 \times 10^4$, orifice height $h = 0{\cdot}04 m$ and exit velocity $u_o = 12\ m/s$. A grid convergence test is carried out with the three grid sizes termed coarse, medium and fine which are $170 \times 146$, $232 \times 166$ and $282 \times 184$ in $x \times y$- directions having a grid refinement factor of $1{\cdot}45$ between the coarse and fine grids. Figure 3 depicts the mean axial velocity profiles at $x/h = 5$ for the three grid sizes. Grid refinement shows successful convergence for those three grid resolutions. The results presented here are obtained by using the fine mesh.

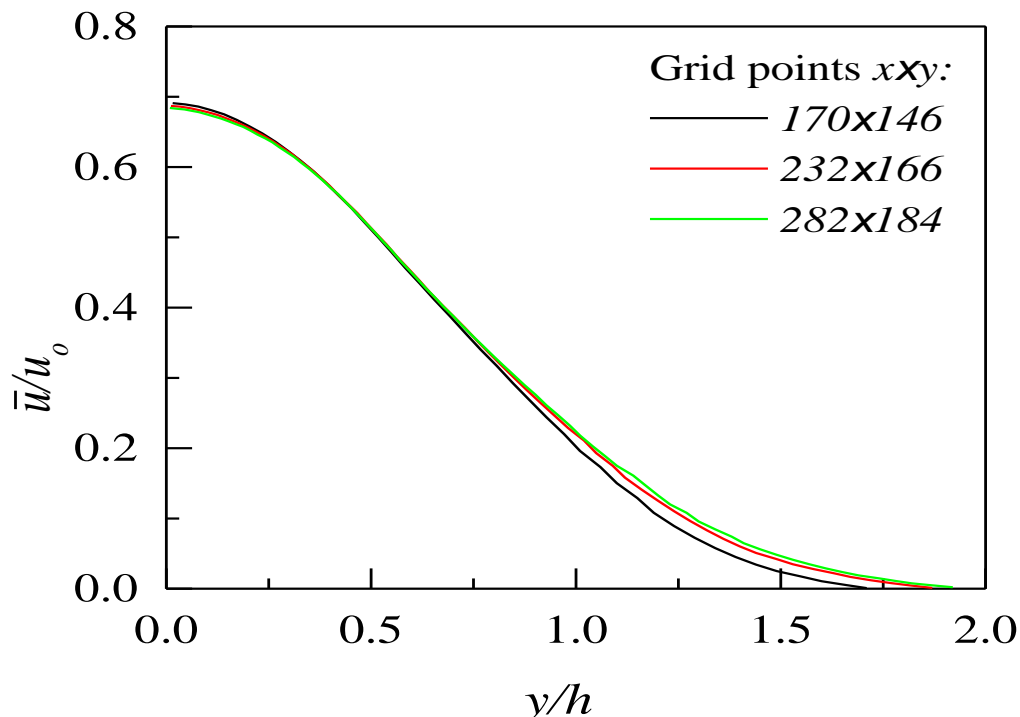

Fig. 3 Axial mean velocity profiles at *x/h=5*.

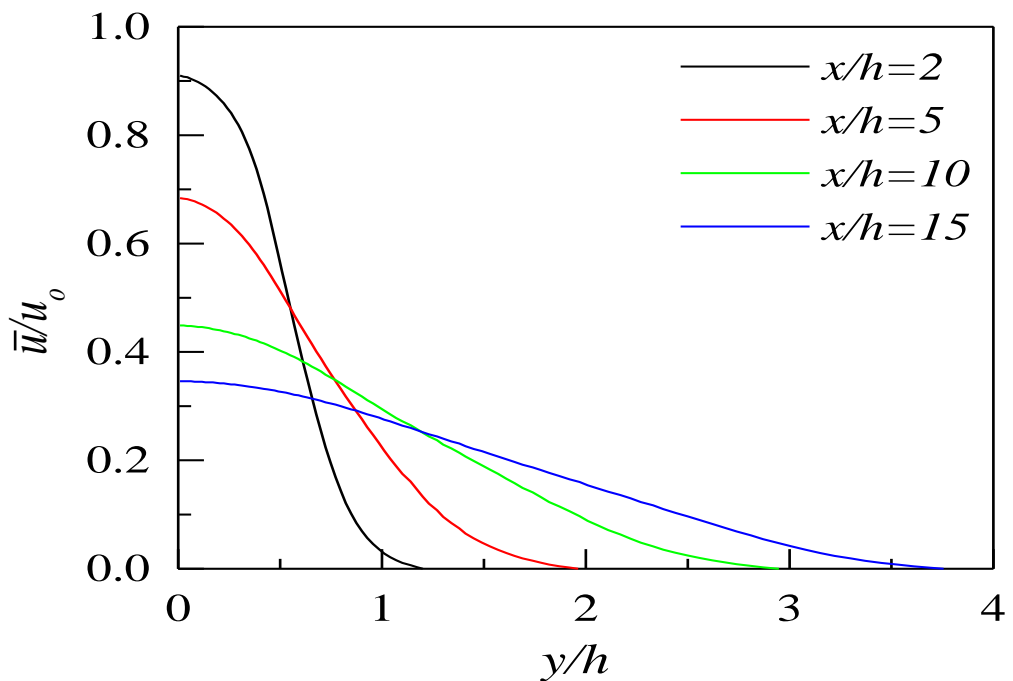

Fig. 4 Axial mean velocity profiles.

## 5 Results and discussion

Closed-form RANS equations (18)-(20), governing the turbulent flow, are solved numerically for a plane jet. The used numerical scheme is found to produce stable solutions for $b_1=3{\cdot}1$ and $b_2=1{\cdot}7$. The values of constants appearing in the functions $f_{ij}(\bar{u})$ for turbulent shear and normal stresses from the stable numerical scheme are $C_s=46$, $C_n=4{\cdot}9$ and $\alpha=1{\cdot}5$ in the initial region of the jet and $C_s=190$, $C_n=8{\cdot}0$ and $\alpha=1$ beyond this region. In determining $f_{12}(\bar{u})$ from Eq. (23c), $\partial u/\partial v$ is replaced by a constant that is absorbed in $C_s$ at $x/h\leq 23$ and by $0.03(x/h+34)$ at $x/h>23$. This section presents the axial velocity, transverse velocity, and turbulent stresses that are extracted in the present simulation and their comparison with the existing theoretical [20], experimental (Exp), and numerical data on planar turbulent jets. Klein *et al*. [21] performed direct numerical simulation (DNS) of a turbulent plane jet with $Re=4\times 10^3$ and compared their results (e.g., mean velocity and Reynolds stresses) with the results of several authors and found them in good agreement. Ramaprian and Chandrasekhara [25] with $Re=1{\cdot}6\times 10^3$ have experimentally investigated the plane jets and presented the profiles of mean velocity and turbulent stresses.

Mean axial velocity $\bar{u}/u_o$ is presented in Fig. 4 against the transverse distance *y/h* at axial locations *x/h=2, 5, 10* and *15*. This figure illustrates that the jet grows with the axial distance due to the entrainment of ambient fluid, and its maximum velocity decreases due to the loss of momentum caused by the interaction with the ambient fluid. The mean velocity profile at *x/h=2* clearly indicates the absence of jet potential core. This is because the equation for Reynolds stresses are algebraic expressions of velocity and pressure, as a consequence, the jet experiences high Reynolds stresses starting from the nozzle exit. Figure 5 presents the mean transverse velocity $\bar{v}/u_o$ against *y/h* at axial locations *x/h=2, 5, 10* and *15*. The figure exhibits that the entrainment of ambient fluid into the jet reduces downstream due to the reduced interaction of the co-flow with the fluid at the jet center.

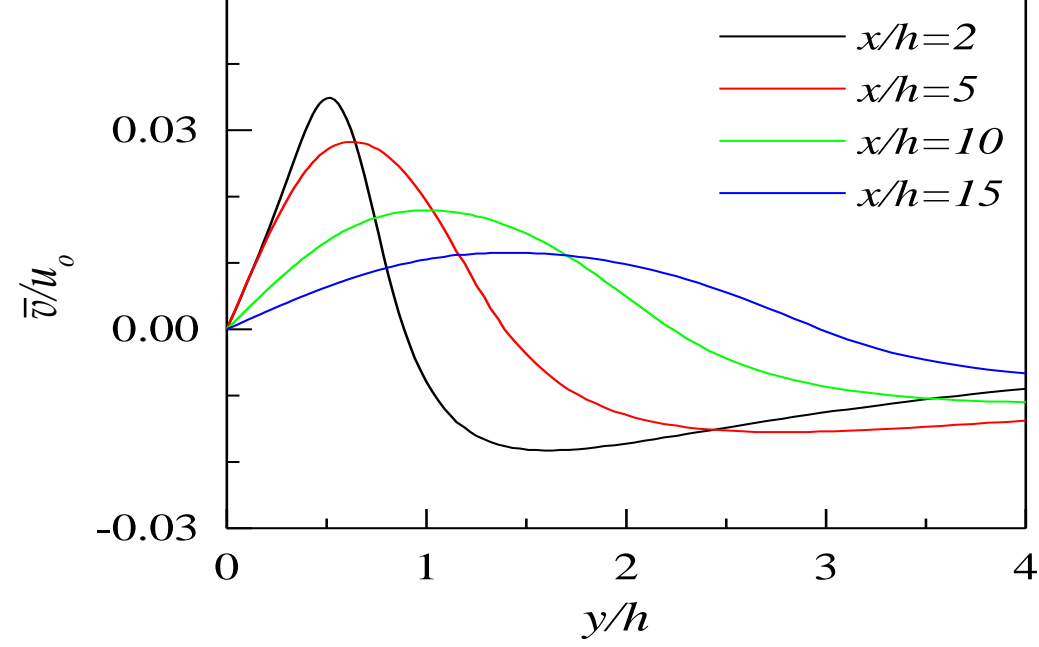

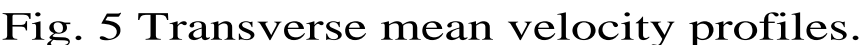
Fig. 5 Transverse mean velocity profiles.

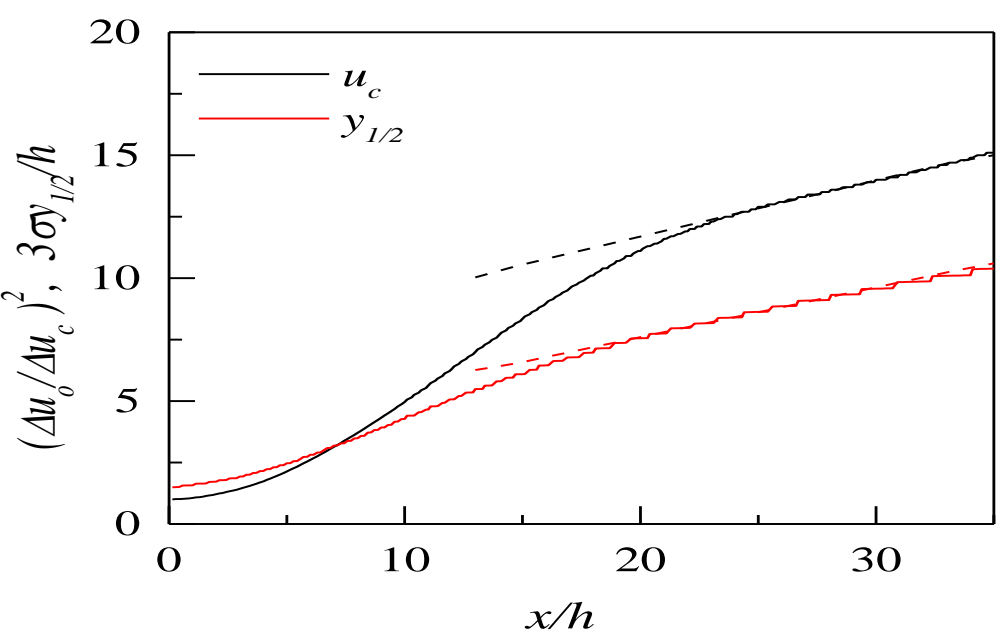

Fig. 6 Decay of centerline velocity and growth of jet width.

Figure 6 shows the growth of half-width $y_{1/2}$ and the decay of centerline velocity $u_c$ of the plane jet with a co-flow (*5%* of the jet exit velocity). Bradbury [26,27] investigated jets with co-flow very extensively. He deduced the relation for jet width using momentum integral equation for the entire range of velocity excess at the jet center $\left(\Delta u_c / u_e\right)$ as

$$\frac{y_{1/2}}{h} = 0.118\frac{x}{h} \Big/ \left(1 + 0.587\sqrt{\frac{x/h}{\lambda(\lambda - 1)}}\right) \qquad (25)$$

where $\lambda = u_o / u_e$. The half-width of the jet is the distance from the jet centerline to the point at which the mean axial velocity $\Delta u = u - u_e$ is half of the centerline velocity $\Delta u_c = u_c - u_e$. The jet growth $\sigma y_{1/2}/h = 0.07(x/h + 17{\cdot}3)$ at $x/h > 18$ (shown as dotted line in the figure) obtained from the present simulation is somewhat at lower rate than the existing values [21,25]. Such a lower rate is likely due to the co-flow [27]. The parameter $\sigma = 1 + 0.0297\sqrt{x/h}$ is assumed from Eq. (25). However, a small co-flow (*0.5%* of the jet exit velocity) is required to overcome the calculation difficulties at the jet outer edge. The decay of centerline velocity $(\Delta u_o/\Delta u_c)^2 = 0{\cdot}23(x/h + 31{\cdot}2)$ appears at $x/h > 23$ (shown as dotted line) is found free from the effect of co-flow where $\Delta u_o = u_o - u_e$. These observed effects of co-flow on the jet growth and centerline velocity decay are consistent with the existing literature [26,27].

Figures 7-8 present normalized mean velocity $\Delta u/\Delta u_c$ against $y/y_{1/2}$ at axial locations $x/h = 2$, $5$, $10$ and $15$ (near field), and at $x/h = 15$, $20$, $25$ and $30$ (far field) along with Gaussian curve [20] of the form $\Delta u/\Delta u_c = exp\left(-0 \cdot 693\eta^2\right)$ at $x/h = 25$ where $\eta = y/y_{1/2}$. Bradbury [26] described the similarity velocity by $\Delta u/\Delta u_c = exp\left[-0 \cdot 6749\eta^2\left(1 + 0.027\eta^2\right)\right]$. However, the present results show that difference between the two velocity profiles due to [20,26] is not discernible even at the jet outer edge. The figures show that results from the present simulation compare well with the Gaussian curve at $x/h \geq 15$ and exhibits self-similarity at all axial locations except at $x/h = 2$.

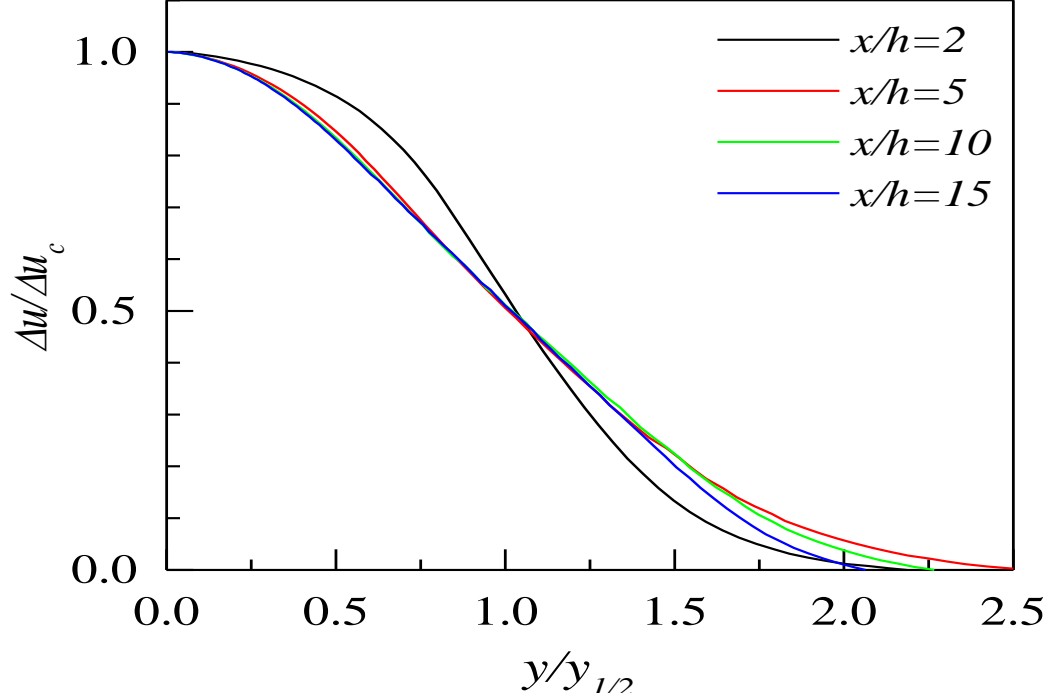

Fig.7 Axial mean velocity profiles.

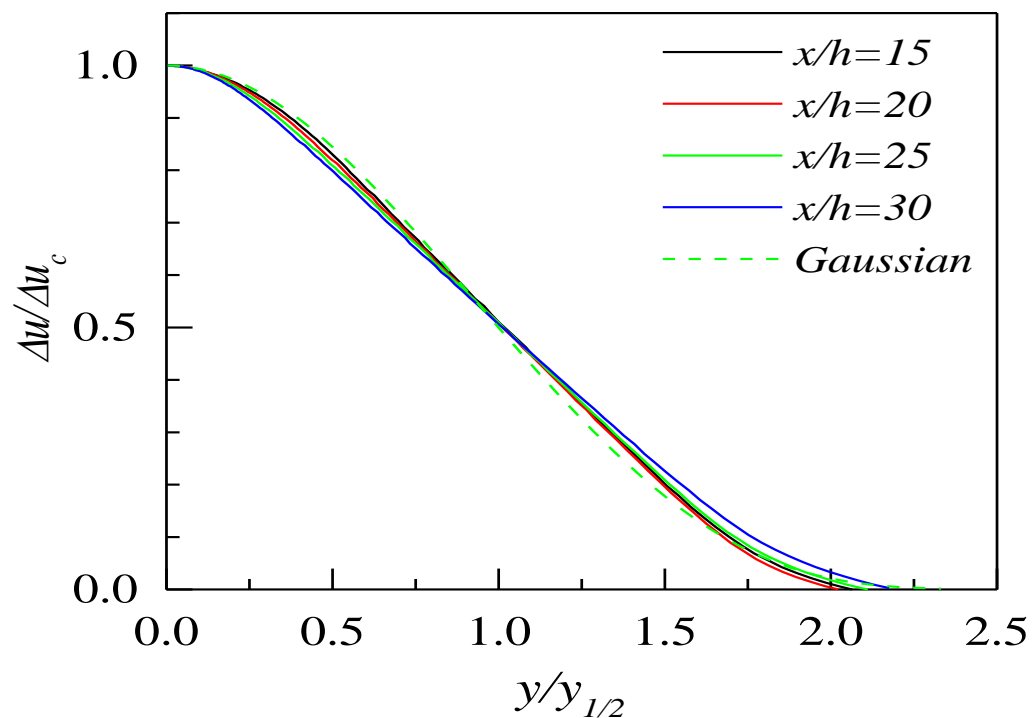

Fig. 8 Axial mean velocity profiles.

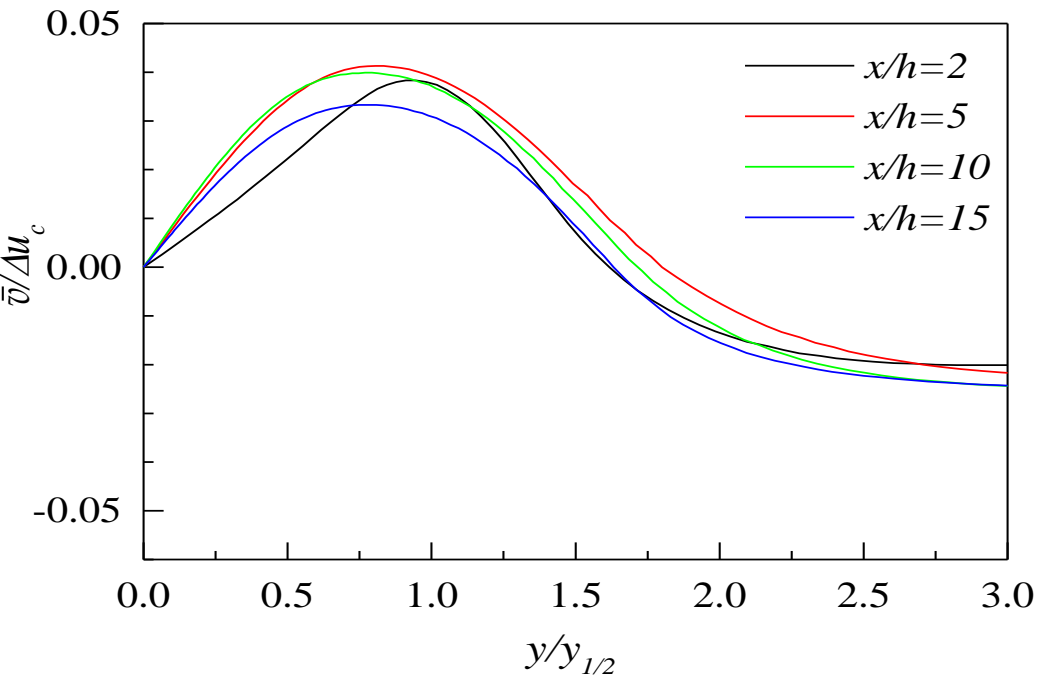

Fig. 9 Transverse mean velocity profiles.

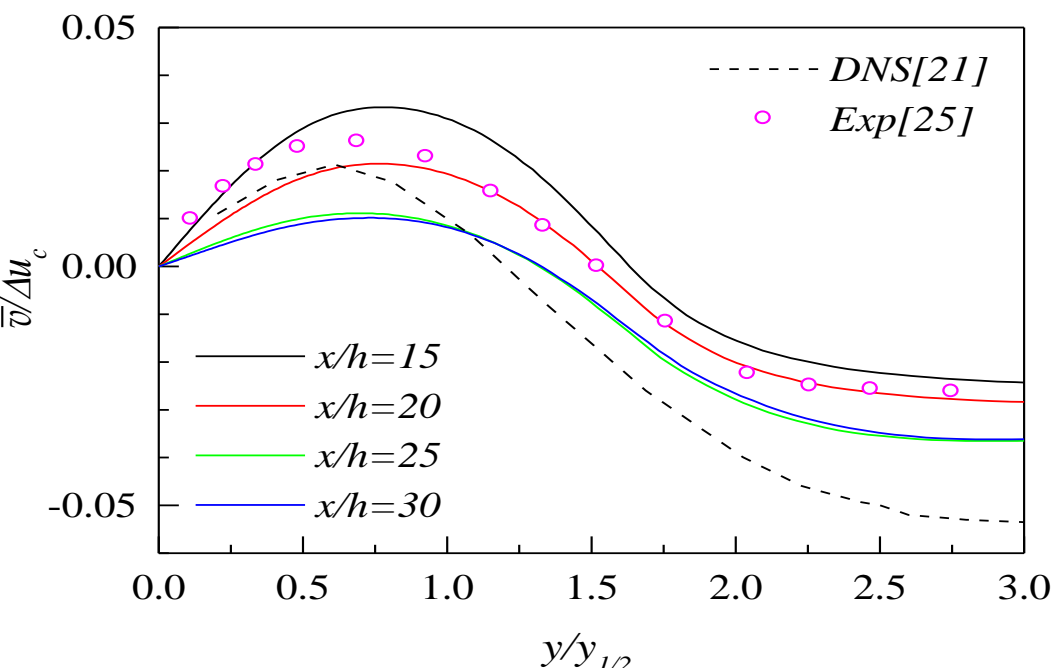

Fig. 10 Transverse mean velocity profiles.

Figures 9-10 depict transverse velocity $\overline{v}/\Delta u_c$ against $y/y_{1/2}$ at axial locations *x/h=2, 5, 10* and *15*, and at *x/h=15, 20, 25* and *30*. The transverse velocity is found to achieve free-stream value at some distance $y/y_{1/2} \geq 2{\cdot}5$. DNS [21] and experimental [25] data for $\overline{v}$-velocity provided in the figure are in satisfactory agreement with the present simulation. The figures show that $\overline{v}$-velocity profiles are self-similar in the far field region at *x/h>20*.

Reynolds normal stresses $\overline{u'u'}/\Delta u_c^2$ and $\overline{v'v'}/\Delta u_c^2$, and shear stress $\overline{u'v'}/\Delta u_c^2$ are calculated from Eq. (24) and plotted against $y/y_{1/2}$ at axial locations *x/h=2, 5, 10* and *15*, and at *x/h=15, 20, 25* and *30*. Figures 11-12 show that the normal stress $\overline{u'u'}$ achieve self-similarity in the far field at *x/h≥20*. Figures 13-14 show that the normal stress $\overline{v'v'}$ achieve self-similarity in the far field like $\overline{u'u'}$. Figures 15-16 show that the shear stress $\overline{u'v'}$ are self-similar in the far field at *x/h≥20*. Reynolds stresses from DNS [21] and measurements [25] added for comparison in Figs. 11-16 are observed in close agreement with those from the present simulation.

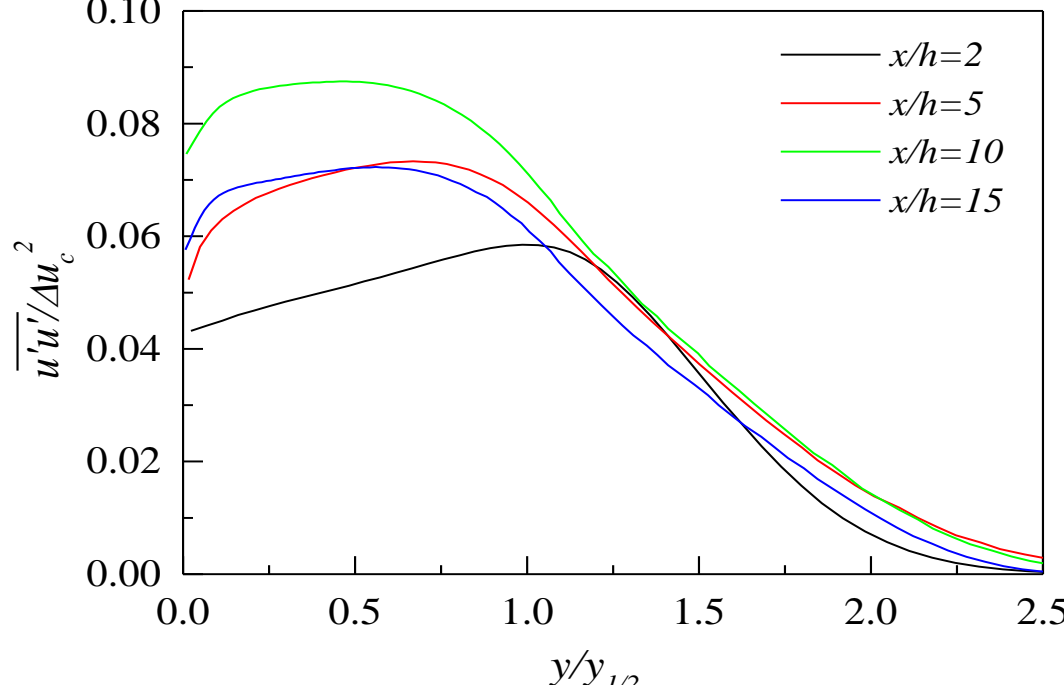

Fig 11. Axial normal stress profiles.

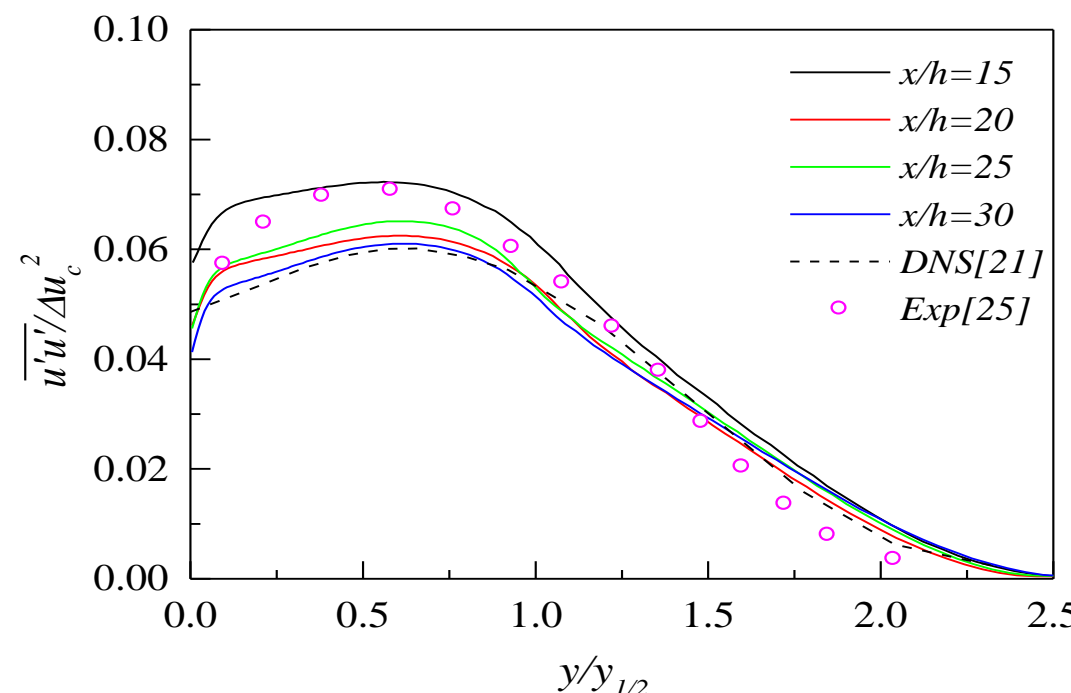

Fig 12. Axial normal stress profiles.

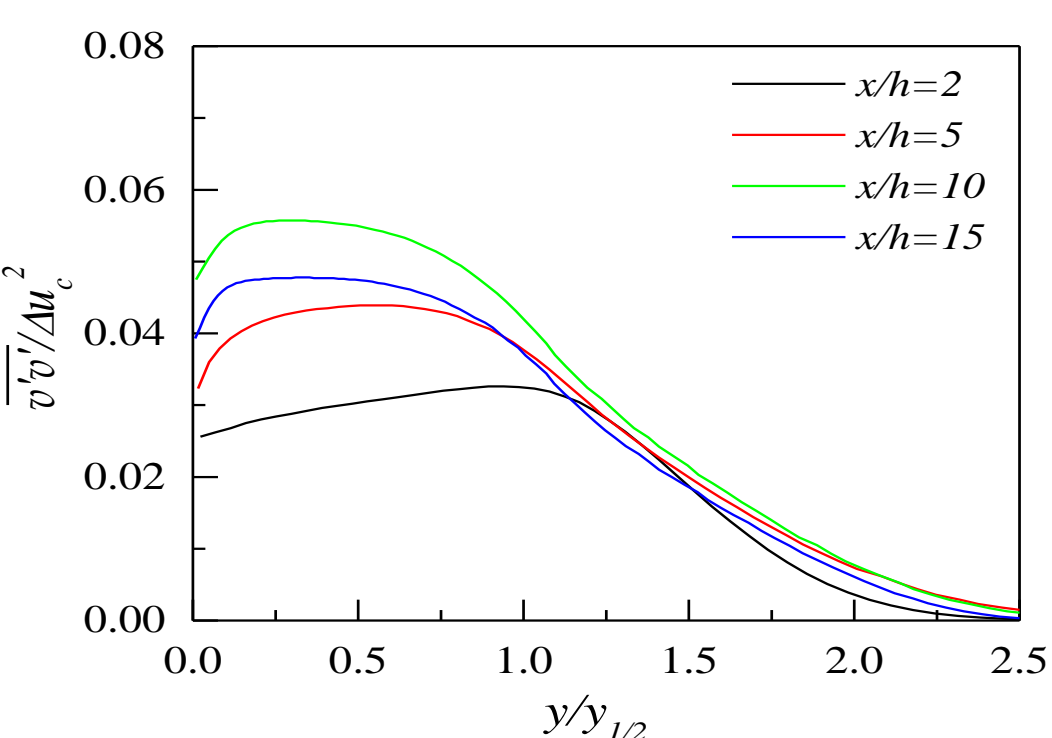

Fig. 13 Transverse normal stress profiles.

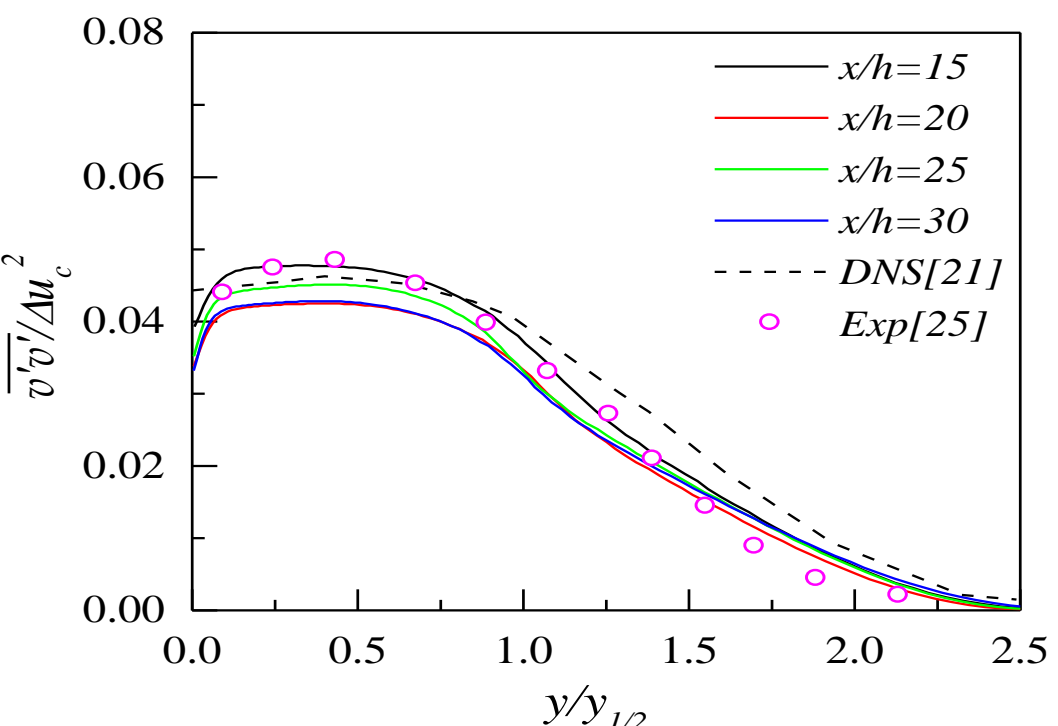

Fig. 14 Transverse normal stress profiles.

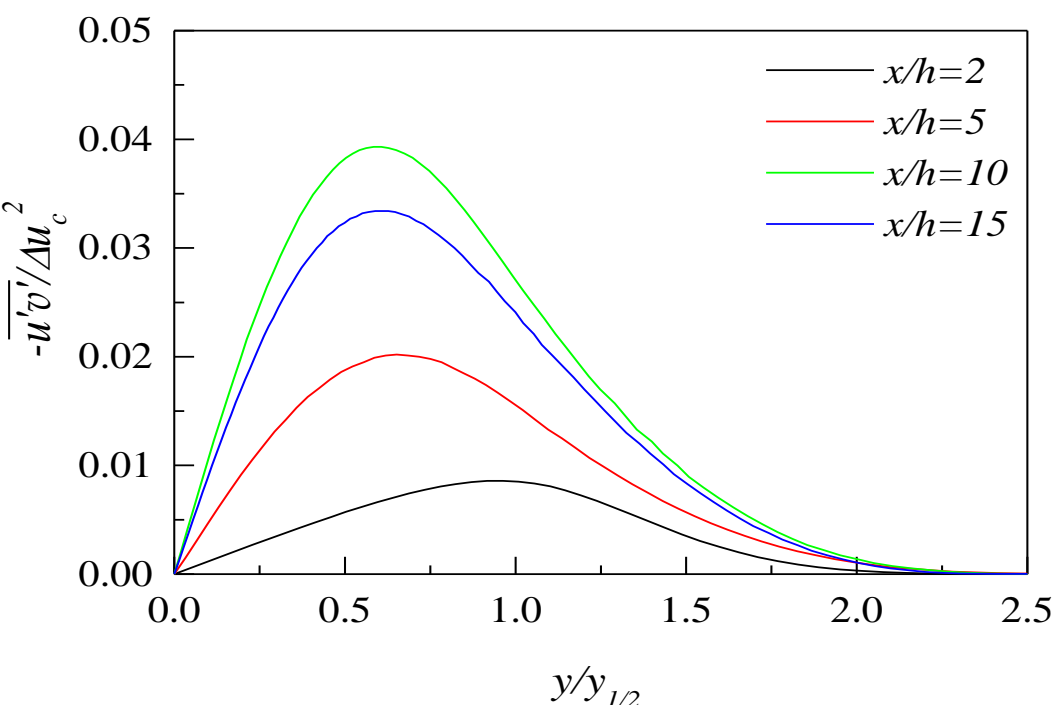

Fig. 15 Reynolds shear stress profiles.

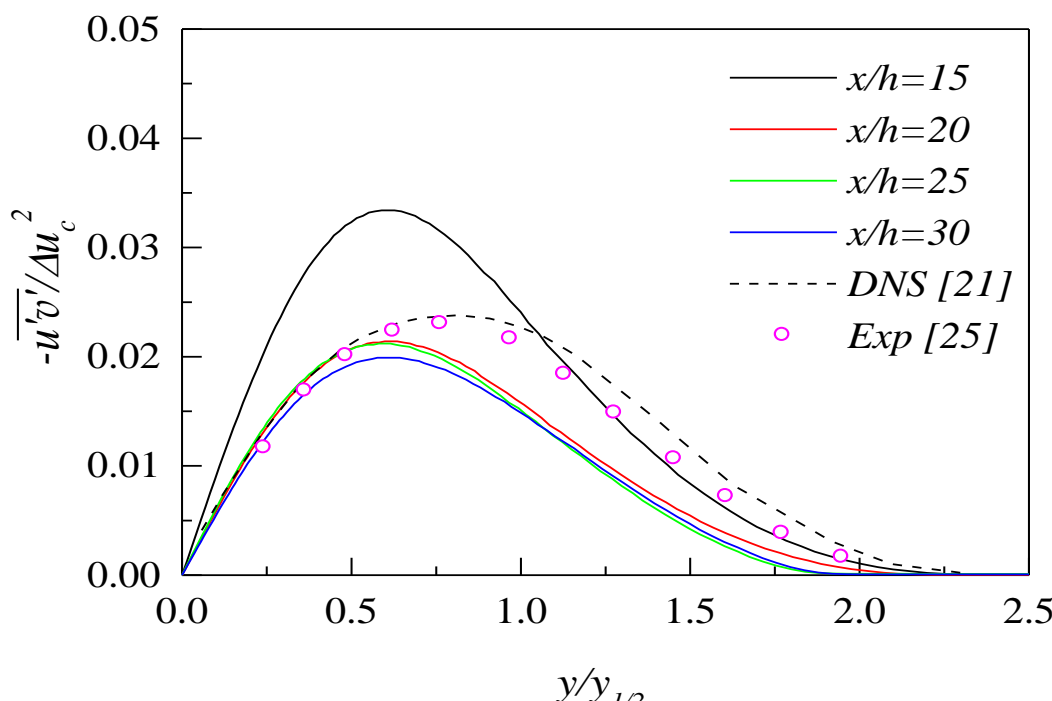

Fig. 16 Reynolds shear stress profiles.

## 6 Conclusion

The new closure equations derived by treating turbulence as CPT are used in the present research to obtain closed-form RANS equations. The numerical solution of those equations for a plane jet provides the mean and turbulent properties of the flow. The inverse square of the centerline velocity and the growth of the shear layer exhibit linear variation with streamwise distance. The scales $u_c$ and $y_{1/2}$ satisfactorily collapse the lateral profiles of mean axial and transverse velocities, as well as turbulent stresses. The simulation results are in good agreement with the existing data, showing the effectiveness of the new closure equations and, hence, confirming the phenomenological relations with the phase transition and restating that laminar-turbulent transition undergoes a continuous phase transition.

## References

1. O. Reynolds, On the dynamical theory of incompressible viscous fluids and determination of the criterion, Phil. Trans. Roy. Soc. A 186, 123-164 (1895).
2. J. Boussinesq, Theorie de l'ecoulement tourbillant, Mem. Pres. Acad. Sci. 23, 46-50 (1877).
3. L. Prandtl, Über die ausgebildete turbulenz, ZAMM 5, 136-139 (1925).
4. A. N. Kolmogorov, Equations of turbulent motion of an incompressible fluid, Izv. Acad. Nauk USSR, Ser. Fiz. 6, 56-58 (1942).
5. L. Prandtl, Über ein neues formelsystem für die ausgebildete turbulenz, Nachr. Akad. Wiss. Göttingen, Math.-Phys.Kl., 6-19 (1945).
6. B. E. Launder and D. B. Spalding, *Mathematical Models of Turbulence* (Academic Press, London, 1972).
7. L. D. Landau, The theory of phase transitions, Nature 138, 840-841 (1936).
8. N. Goldenfeld, Roughness induced critical phenomena in a turbulent flow, Phys. Rev. Lett. 96, 0445031-4 (2006).
9. S. T. Bramwell, P. C. W. Holdsworth, and J.-F. Pinton, Universality of rare fluctuations in turbulence and critical phenomena, Nature 396, 552-554 (1998).
10. Ch. Chatelain, P. E. Berche, and B. Berche, Second-order phase transition induced by deterministic fluctuations in aperiodic eight-state Potts models, Eur. Phys. J. B 7, 439-449 (1999).
11. W. S. Gan, Application of spontaneous broken symmetry to turbulence, Proc. ICSV16, Krakow, Poland, 585-589 (2009).
12. K. Avila, D. Moxey, A. de Lozar, M. Avila, D. Barkley, and B. Hof, The onset of turbulence in pipe flow, Science 333, 192-196 (2011).
13. N. Vladimirova, S. Derevyanko, and G. Falkovich, Phase transitions in wave turbulence, Phys. Rev. E 85, 010101-4 (2012).
14. L. Shi, G. Lemoult, K. Avila, S. Jalikop, M. Avila, and B. Hof, The universality class of the transition to turbulence, arXiv:1504.03304, physics.flu-dyn (2015).
15. N. Goldenfeld and H.-Y. Shih, Turbulence as a problem in nonequilibrium statistical mechanics, J. Stat. Phys. 167, 575-594 (2017).
16. M. E. Fisher, Correlation functions and the critical region of simple fluids, J. Math. Phys. 5, 944-962 (1964).

17. M. J. Klein and L. Tisza, Theory of critical fluctuations, Technical Report No. III, ERL, Massachusetts, 1-20 (1949).
18. T. Vicsek and A. Zafeiris, Collective motion, Phys. Reports 517, 71-140 (2012).
19. P. A. Durbin and B. A. P. Reif, *Statistical Theory and Modeling for Turbulent Flows* (John Wiley and Sons, 2011).
20. R. C. Deo, J. Mi, and G. J. Nathan, The influence of Reynolds number on a plane jet, Phys. Fluids 20, 075108:1-16 (2008).
21. M. Klein, A. Sadiki, and J. Janicka, Investigation of the influence of the Reynolds number on a plane jet using direct numerical simulation, Int. J. Heat and Fluid Flow 24, 785-794 (2003).
22. M. A. Azim, Numerical study of closely spaced twin circular jets, Int. J. Fluid Mech. Res. 47, 407-418 (2020).
23. S. V. Patankar and D. B. Spalding, A calculation procedure for heat, mass and momentum transfer in three dimensional parabolic flows, Int. J. Heat Mass Transfer 15, 1787-1806 (1972).
24. L. H. Thomas, Elliptic problems in linear difference equations over a network, Watson Sci. Comput. Lab. Report, Columbia University, New York (1949).
25. B. R. Ramaprian and M. S. Chandrasekhara, LDA measurements in plane turbulent jets, ASME J. Fluids Eng. 107, 264-271 (1985).
26. L. J. S. Bradbury, The structure of a self-preserving turbulent jet, J. Fluid Mech. 23, 31-64 (1965).
27. L. J. S. Bradbury, Simple expressions for the spread of turbulent jets, Aero. Qtr. 18, 133-142 (1967).

**Comments:** In this version, i) the implicit function is derived systematically in Sec. 2, ii) the axial velocity $\bar{u}$ is clearly used as the order parameter instead of $\bar{u}_i$ in Sec. 2, iii) the transverse coordinates, $y$ and $y'$, are identified in Sec. 2 for mean and turbulent quantities, and iv) the results are presented for 75% of the computational domain.